# XP2021 Experience Report:
# Five Strategies for the Future of Work: Accelerating Innovation through Tech Transfer

STEVEN FRASER, Innoxec, Santa Clara, CA. USA

Abstract: This experience report outlines five tech transfer strategies developed over a period of 25 years at four Global 1000 companies (HP, Cisco, Qualcomm, and Nortel) to mitigate R&D challenges associated with duplicated effort, product quality, and time-to-market. The five strategies accelerate innovation through open knowledge sharing, rather than licensing intellectual property rights (IPR) such as patents, trade secrets, and copyrights. The strategies are based on corporate tech forums, conference panels, exploratory workshops, research reviews (at universities and companies), and talent exchanges. While the initial objective was to foster the corporate adoption of software best practices, over time the strategies had broader impact on company innovation, including incubating cross-company R&D collaborations, capturing organizational memory, cultivating and leveraging external research partnerships, and feeding company talent pipelines.

## 1. INTRODUCTION

This report relates my corporate tech transfer experiences as distilled into five strategies. In the context of this paper, tech transfer describes the process where value is demonstrated and disseminated via sharing knowledge from its originators (internal or external to an organization) to a broader company, university, and/or public community, and not by the licensing of tangible intellectual property rights (IPR). Fig. 1 illustrates a typical funnel of commercialization for a company. My tech transfer strategies increased my company's capacity to share information within, into, and out of the company. The skills I used to catalyze tech transfer were of a general technical nature, complemented by my acquired business and people skills as summarized in Table 1. My initial advice to individuals and teams leading tech transfer programs is to think more broadly about people and the business beyond a narrow focus on technology, to leverage opportunities, and to exploit serendipity.

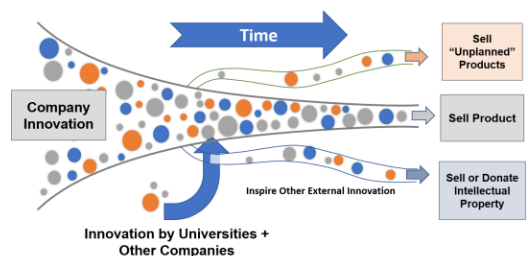

Figure 1. Tech (Knowledge) Transfer and the Porous "Funnel of Open Innovation" [1]

| Tech Community Skills | Business Skills | People Skills |
|---|---|---|
| Agile + Waterfall Processes | Brand Management | Authoring + Teaching |
| Analysis + Design + Testing | Budgeting + Logistics | Collaborating + Learning |
| Educating + Presenting | Business Development | Diplomacy + Politics |
| Governance + Standards | Compliance + Ethics | Facilitating + Influencing |
| Networking + Publishing | Event Management | Leadership + Curiosity |
| Open Innovation + Invention | Fundraising + Philanthropy | Mentoring + Coaching |
| Research Portfolio Management | Legal + Contracts | Negotiating + Persisting |
| Scientific Method + Engineering | Program Management | Public Speaking + Writing |
| Software + Hardware + Systems | Sales + Marketing | Recruiting + Outsourcing |
| University-Company Relations | Strategy + Planning | Stakeholder Alignment |

Table 1. Skills Used and Acquired Catalyzing Tech Transfer

Overall, my approach had three parts: share knowledge, foster a culture that promotes learning, and increase the visibility of internal company "best practices." I discovered that shared learning harnesses employee enthusiasm, helps connect experts, and fosters talent retention. My personal organizational contexts are summarized in Table 2. My tech transfer ecosystems took many forms. Our broadest undertaking was the orchestration of multiple company-wide proprietary internal tech forums. While initially motivated by the need to catalyze and leverage the adoption of software best practices, our goals broadened to include tech transfer through collaborations with university researchers. I refined and applied my tech transfer strategies as I worked for large Global 1000 companies (Nortel, Qualcomm, Cisco, and HP), start-ups, and universities. The shared learning ecosystems I created were designed to incubate company collaborations and to accelerate innovation. This report paints a broad picture focused on a holistic view of tech transfer that should appeal to practitioners, product managers, executives, and researchers. The five shared learning strategies I will explore in order are: global corporate technology forums, conference panels, exploratory workshops, research review meetings (both university and company), and talent exchanges. Note that while "Agile Practices" were one of the technologies for which I catalyzed adoption, the tech transfer strategies I developed and applied were technology agnostic and have been applied to the adoption of software, hardware, systems, and applications.

|  | BNR/Nortel | CMU's SEI | Qualcomm | Cisco | HP | Innoxec |
|---|---|---|---|---|---|---|
| Author's Role | Researcher, Manager—Process Engineering, Senior Manager—External Research | Visiting Scientist (Nortel Consultant) | Senior Staff, Qualcomm Learning Center | Director, Cisco Research Center | Lead, HP Global University Programs | Director, Advisory Services |
| Organization Focus | High Availability Telecom Systems | Improved Software Maturity through Best-Practices | Mobility (Wireless) System Technologies | Collaboration and Communication Systems | Printer and Personal Computing Systems | Advising Companies on Tech Transfer Strategies |
| Tech Transfer Focus | Software Best Practices, External Research | Domain Engineering (software reuse) | Agile Software Best Practices, Wireless Technology | External Research, Ph.D. Recruiting, Employee Upskilling | HP University Partnerships | Advisory Services for Tech Transfer and Best Practices |
| Location | Ottawa, Canada | Pittsburgh, USA | San Diego, USA | San Jose, USA | Palo Alto, USA | Santa Clara, USA |
| Development Community | ~25k Engineers >10 global R&D labs | Government Contractors | ~10k Engineers ~ 5 R&D labs | ~25k Engineers >10 R&D labs | 20k Engineers > 15 R&D labs | Varies by contract |
| Process | Waterfall | Waterfall | Early Agile Transition | Broad Agile Transition | Broad Agile Transition | Varies by Contract |

Table 2. Author's Organizational Contexts

2. GLOBAL CORPORATE TECH FORUMS: ACCELERATING TECH TRANSFER THROUGH SHARED LEARNING

In the early 1990s during the genesis of what would become known as agile practices, my mission was to foster the adoption of "software reuse." The intent was to identify system "commonality" and minimize system "variability," i.e., limiting multiple versions of similar but slightly different code. Initially, my strategy was to use indexing tools to build a searchable software component catalog, since if teams could find the software required, they would have no need to duplicate effort and code volume. Unfortunately, my initial reuse tool developed for the exploratory "Telos" project at BNR/Nortel in Smalltalk-80 was slow, both for indexing and retrieval [2]. Looking to scale reuse indexing to Nortel's flagship product software library (at that time ~10-15 MLOC per release), I discovered a Nortel support tool called CD-ROM Protel. The tool provided an indexed snapshot of the code base once every three months and was intended for product support teams. The tool was of limited value for developers, due to indexing and snail-mail distribution delays.

Through my interviews and partnerships with engineers across BNR/Nortel's global engineering community I learned that there were many departmental "best practices" and key learnings that went unshared across the company. In several cases, I identified situations where our competitors knew more about Nortel technology and process advances through ACM or IEEE publications by Nortel experts, than did our broad employee community. It was obvious to me that the company lacked an effective internal sharing mechanism across business units. At this time, because of increased demand for my personal consulting and facilitation services, I realized that my workshops and tech transfer consulting would not scale up.

While my focus was software reuse, on my own initiative I decided to take a broader approach and organize a corporate wide "Design Forum." My reasoning was that a "Software Reuse Forum" would have limited influence and appeal. My goal was to build a collaborative ecosystem to motivate engineers and senior managers across the company to participate, share design best practices, and foster a "learning community" to tech transfer practices across business units. I achieved this by leveraging what I now recognize as "influence strategies" (Table 3) [3,4]. For example, it was important to obtain visible sponsorship from executives (leveraging their "authority") to encourage participation. My forums consisted of multi-day technical programs featuring employee presentations, solicited through open calls, curated by peer reviews, and sponsored by the CTO and other company executives. Key elements of my tech forums included:
- *Bootstrapping forums by inviting company staff with ACM/IEEE publications to submit previously published papers and adding proprietary details not included in public presentations*
- *Using a double-blind peer review process to select papers based on content relevance and novelty (and to provide feedback to authors while avoiding the influences of author seniority during the review process)*
- *A modular program with tracks specific to software, hardware, and system best practices (to focus presentation based on attendee interests)*
- *Keynotes to increase forum visibility and attract a broad spectrum of participation*
- *Coaching assistance for presenters to improve presentation skills*
- *Executive hosts for keynote speakers—a dual influence strategy to catalyze executive "connections" through proximity and to increase internal visibility and demonstrate company sponsorship for sharing best practices*
- *Video conferencing (e.g., PictureTel, VTel, WebEx, TelePresence) to enable the idea of a "low cost, no travel" down-the-hall or desktop company conference*
- *Sponsorship by senior executives to "give permission" for participation*
- *Session recordings for attendees with schedule conflicts—initially video tapes, and recently video-on-demand*
- *Conference proceedings (initially in print, later digitally) for organizational memory—at Qualcomm, the proceedings were used to catalyze a company technical journal*

Executives were attracted to the idea of fostering increased collaboration across business units and inspired innovation. Once the forum was established and successful, it catalyzed staff engagement. This in turn increased internal competition to participate. Available presentations slots became "scarce" and highly sought as a form of recognition. "Best Paper Awards" and customized conference tchotchkes (swag) provided additional incentives for participation, both by presenters and attendees. In the days of video conferencing, it was useful to leverage the influence of "proximity" and "snacks" when engineers from an R&D Lab would meet serendipitously in their local corporate video conference room at one of more than 25 labs.

Many keynote speakers participated from BNR/Nortel's lab in Silicon Valley even though the conference was produced from our Ottawa headquarters. A virtual conference made it possible to include prominent software keynote speakers on our Tech Forum program, independent of their physical location.

| |
|---|
| • Authority (power relationship) |
| • Collaboration (multiplies capabilities) |
| • Commitment (agreement) |
| • Consistency (go with the flow) |
| • Fear (do it or you'll be sorry!) |
| • Inspiration (go with a good idea) |
| • Liking (follow a friend) |
| • Proximity (local insight, food & drink) |
| • Reciprocity (give a gift back) |
| • Reward (promise a "payoff") |
| • Scarcity (get before it disappears) |
| • Social Proof (follow the crowd) |

Table 3. Influence Strategies [3,4]

Following a "Call for Participation" (Fig. 2), I collected submissions and coordinated reviews by email. However, even with 30 papers and a similar number of reviewers, I almost lost my sanity coordinating details (with 3 or more reviews per submission). For the second iteration of the BNR/Nortel Design Forum, I developed software to manage submissions and reviews. With Qualcomm's QTech and Cisco's CTech I used third-party conference tools.

Once submissions were reviewed (with three to five reviews per paper) and I had confirmed keynotes and executive hosts, the program schedule was developed in collaboration with my Vice-Chair and program committee, accounting for factors such as presentation topic and presenter location. If presenters were in India, their presentations had to be early in the program day given that the Forum started between 8 am and 9 am on the East Coast of

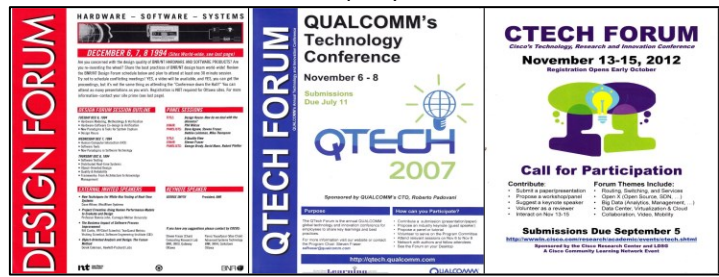

Figure 2. Sample "Call for Participation" Posters

North America which was after regular business hours in India. Another consideration was to ensure program diversity. For example, if a dozen presentations were highly rated, this did not imply that all would appear on the final forum program since at most two sessions (a total of six presentations) would normally be devoted to a single topic.

For Q&A sessions, it was challenging to solicit questions from our global audience, so as part of pre-forum preparations, I required presenters to provide two questions. However, much to our surprise, some presenters were not prepared for their own questions, which led to some embarrassing silences. Generally having prepared questions was useful to kick-start Q&A sessions. As our process grew more mature, I noticed that presenters might attend only their presentation slot (and not their entire 90-minute session). To promote dialogue between session presenters, I adapted a program format to run "Q&A" panel style, with three 20-minute presentations followed by 30 minutes of discussion amongst the presenters and audience. Another discovery was that presentation titles were important, and a bit of drama or humor (as with newspaper headlines) improved audience interest. Another recommendation to presenters was to reduce both the complexity and number of slides to improve readability.

Presentations were based on "papers." However, to allay the fears of middle management (too much time writing papers instead of code), presenters were encouraged to reuse content from internal reports or previously published papers from ACM/IEEE Conferences. If the papers had appeared at a public conference, presenters were encouraged to add a proprietary story-behind-the-public-story to add value for the internal presentation. Self-plagiarism was encouraged if the material had not been presented at a previous Forum.

The proprietary nature of presentations necessitated some additional precautions. If the content was subject to US government export compliance regulations such as ITAR (International Traffic in Arms Regulations) and EAR (Export Administration Regulations), then registrants needed to be vetted and screened to bar the presence of unlicensed staff from "Group D" or "Group E" countries such as China, Cuba, Iran, Iraq, Kazakhstan, North Korea, Russia, etc. Coordinating the security of sessions was particularly challenging given the global nature of the conference. Promotion was also challenging, and we used a mix of print (Fig. 2) and online marketing.

Virtual forums proved useful to share best practices within multiple corporations, including BNR/Nortel, Qualcomm, and Cisco. Company-personalized stories of technology adoption were more compelling than those made by third party short course instructors and increased the "stickiness" of learning. That said, keynote presentations by software thought leaders such as Marc Andreessen, Kent Beck, Barry Boehm, Fred Brooks, Grady Booch, Ward Cunningham, Whitfield Diffie, James Gosling, Jim Gray, Bill Joy, Guy Kawasaki, Norm Kerth, Steve McConnell, Dave Parnas, Linda Rising, Douglas Schmidt, Mary Shaw, Don Tapscott, and Ed Yourdon, to name but a few of our keynotes, increased visibility for best practices and fostered relationships between speakers and staff. Forums also proved a ready-made venue to showcase research collaborations with university researchers. The connections made through forums helped incubate and sustain research relationships, often extending to talent exchanges (internships, sabbaticals, and full-time hires).

Once a regular cadence of forums was established, managers would frequently remind their staff of the next forum to prepare presentations to create visibility for their teams and to share best practices. The forum also proved attractive to new grad hires as a vehicle to share their doctoral research beyond their local teams and as collateral for recruiting, e.g., at Nortel Santa Clara Lab a Gallery of Forum posters were displayed prominently in the entry hall. Another observation relates to the programmatic nature of shared learning programs. They have a beginning and an end. My personal practice is to celebrate the conclusion of a tech forum with a "cast party" and recognition for my colleagues who have collaboratively delivered the program.

In terms of submission numbers, regional labs were exceptionally well represented, disproportionately more so than their R&D population would suggest. Staff at recently acquired companies were invited to participate and contribute as part of their onboarding process such as formerly "external" staff from XROS at Nortel, Flarion at Qualcomm, and Tandberg at Cisco. At Qualcomm, the tech forum was expanded to include an internal technology trade show and the QTech Forum persisted for 10+ years following my departure for Cisco. I should also note my disappointment, that on the day I started at HP in 2016, the HP corporate wide tech forum that inspired my Design Forum, QTech Forum, and CTech Forum was cancelled due to cost constraints.

3. CONFERENCE PANELS: INSPIRING INNOVATION THROUGH EXPERT DISCUSSIONS AND DEBATES

A corporate technology conference benefits from "external" influencers and stakeholders that inspire new directions, technologies, and evangelize "best practices." To that end, scouting university consortia and research programs has proved a useful source to me of ideas for keynotes, topics, and new grad talent. As a researcher at BNR/Nortel, I discovered that it could be challenging to get approval to present at external conferences due to a complex publication "release" process that took several months. The process required sign-off by layers of engineering management, patents, public relations, marketing, sales, and senior executive management. When I had to withdraw my name as a publication co-author with a professor from the Naval Post Graduate School (NPGS) due to approval delays, I sought new catalysts (beyond research papers) to give me collateral to interact with key thought leaders at conferences.

One of my most successful strategies was to act as a conference panel "impresario." Traditionally, an impresario organizes operas, concerts, and other entertainment. My challenge was to organize discussions by industry and academic thought leaders. To date, I have organized 100+ panels and some of my favorites are [5,6,7,8,9,10]. My guiding principles for panels include:

- *Design panels around interesting and topical subjects featuring diverse and often contrarian panelists from both industry and academia*
- *Publish a panel abstract together with bios and position statements to save time with introductions*
- *As panel facilitator (impresario) do not participate as a panelist (otherwise you run the risk of using your facilitation role to add weight to your perspectives to the detriment of other views)*
- *Avoid "mini" presentations and allow only brief opening statements by panelists, since the objective is discussion and debating perspectives, rather than extended evangelization or talking "at" fellow panelists*
- *As facilitator seek out interesting questions in advance to catalyze new directions in discussion*
- *When taking questions, it is prudent to take them in writing (chat for virtual sessions) to keep discussion flowing and to avoid rambling participant pontifications or lengthy queues of audience members asking questions no longer relevant*
- *Poll the audience frequently to inspire new conversations and sustain engagement*
- *Wrap up the discussion on time with short (1 minute) summary statements from the panelists*
- *To "tech transfer" and evangelize outcomes beyond the memories of those present, document the discussion and publish a summary report in post-conference proceedings or a popular journal*

One challenge with panel descriptions is that they are an edited collection of panelist bios and position statements setting the context for an "event" rather than a collaboratively written paper. This makes it difficult for panelists, particularly at competing companies, to sign copyright releases designed for paper collaborators.

4. EXPLORATORY WORKSHOPS: IDENTIFYING TEAM NORMS, MENTAL MODELS, AND VOCABULARY

There are two forms of workshops. The first, commonly found co-located with international public conferences, is essentially a mini conference consisting of a collection of presentations, panels, and keynotes. A second format consists of a facilitated discussion. The participants explore a subject area by identifying shared vocabulary, norms, and mental models, or they build a roadmap or vision for future work by collecting, categorizing, analyzing, or discussing topics of mutual interest. Through the work of Clifford Saunders and John Warfield as described in my collaboration with Cliff [11], I developed an exploratory workshop process to help teams identify opportunities for software reuse. I later generalized this process to a team-based domain engineering workshop.

Workshops consisted of developing a "trigger question" in partnership with executive or management workshop sponsors and using this question to focus discussion. A variety of facilitation techniques were used including Nominal Group Techniques (NGT) to brainstorm ideas, categorization (to group ideas), and Interpretive Structural Modeling (ISM) to order categories of ideas based on the original trigger question. Another technique used a "visioning" process.

These workshops proved useful in identifying departmental duplications and in one case avoided $22M in contractual penalties. As a result of presenting [11] at ICSR'93, my workshop strategy brought me to the attention of Carnegie Mellon University's Software Engineering Institute (SEI) in Pittsburgh, where I was invited to collaborate for one year as a Visiting Scientist. With my year in Pittsburgh, I acquired many valuable insights both in the development of software best practices and tech transfer strategies. I learned new workshop facilitation techniques, learning/training strategies, and expanded my academic and company networks.

Interactive facilitated exploratory workshops also proved useful at conferences. At OOPSLA'94 I organized a workshop titled: *How Do Teams Shape Objects? How do Objects Shape Teams*? The workshop attracted Kent Beck, Jim Coplien, Ward Cunningham, Norm Kerth, Linda Rising, and others. OOPLSA'94 also featured the first of my many conference panels. Since then, I have continued to organize workshops at ACM's OOPSLA/SPLASH Conference and at the Agile Alliance's XP Conference, often in partnership with my colleague Dennis Mancl. As described earlier, panels for individuals constrained by corporate policies often offer a lower barrier to participation in terms of "paper release" processes (at least as a panel chair).

Other exploratory workshops, such as the Dagstuhl Seminar on Technical Debt I attended in April 2016 [12], have proved instrumental in expanding my personal network of academics and industry practitioners and to catalyze new connections for HP Engineers.

5. RESEARCH REVIEWS: SHARING AND PRIORITIZING RESEARCH RESULTS

University consortia and departments with industry partnership programs frequently host research reviews, consortia sponsor meetings, or technical conferences. These meetings report on progress, solicit feedback, reprioritize activities, and catalyze opportunities for interaction between industry partners, academics, and students. New relationships incubated at these meetings led to new research programs, scholarships, sabbaticals, and student hires. As a tech transfer change agent, I recognized that review meetings complemented panels and conferences as a key component in my strategy to connect external researchers with company engineers and to scout emergent technologies.

In the days of face-to-face conferences and meetings, it was challenging for attendees to schedule and budget for travel. At Cisco and HP, I curated a program of suggested university review meetings for company engineers and executives to coordinate and schedule participation. With the onset of the COVID-19 pandemic, most conference and university review meetings became virtual, reaching a far broader audience. Costs (registration and associated travel) have been reduced while flexibility has increased. For example, I have attended Stanford, Berkeley, and MIT events on the same day. There is no need to budget or schedule trans-continental travel. My current (Spring 2021) recommendations for these virtual forums include UCLA's ECE's Virtual Research Review, USC's Center for Systems and Software Engineering, Stanford's HAI (Human Centered AI Institute) Conference, and MIT's Media Lab sponsor meetings.

At Cisco, I developed a "Research Commons" for sharing internal research and mentoring new grad doctoral hires. Modeled on consortia meetings, Research Commons were a series of quarterly presentations and discussions. This forum provided an informative forum for both early-career full time research staff and interns. These proved useful to connect staff to mentors, share interests, and to review Cisco's research and engineering strategies. At HP, as part of our global university program strategy roll-out, I created an internal virtual community supported by Microsoft's enterprise Yammer social networking tool. Meetings connected global participants from HP Labs in North America, South America, Europe, and Asia/Pacific and focused on sharing best practices for university research interactions and recruiting. In summary, collaborative "discussions" and "reviews" complemented by corporate "Grand Challenges" and "Request for Proposal" (RFP) programs are effective catalysts for tech transfer as I have described previously in [13,14,15,16].

6. TALENT EXCHANGES: PROMOTING SHARED LEARNING AND INNOVATION

Throughout my career, I have orchestrated talent exchanges through sabbaticals, fellowships, and visiting scientistships, catalyzing connections made through my participation in university visits, research reviews, and public ACM, IEEE, and XP conferences. These connections improved my ability to scout for emergent technologies, to build research partnerships to catalyze tech transfer, and to recruit doctoral and post-doctoral talent from university research labs.

More recently, I have appreciated the need for mentorship and internship programs and regretted that I had not been able to leverage such programs during my masters (Queen's) and doctoral (McGill) studies. A combination of corporate funding and interest is key to enabling such programs combined with company scholarships, fellowships, sabbaticals, adjunct professorships, and visiting scientistships. The injection of interns into an organization brings new ideas unconstrained by the norms, beliefs, and values of the organization. Some care must be taken to ensure that proprietary trade secrets are not inadvertently disclosed, and precautions are required to observe regulations related to "export compliance." That said, "onboarding" for new staff, particularly interns, should catalyze ongoing collaborations and shared interests to encourage the return of interns as full-time staff upon graduation. Care should be taken to understand that doctoral interns have significantly more experience than undergrads and are motivated by specific career directions, including academia.

My other learning programs leveraged external interactions with universities and companies. For example, I managed early BNR employee participation in Ontario's Consortium for Graduate Education in Software Engineering (ConGESE) Master's program and led the "Software Best Practice" program at Qualcomm. My Qualcomm program featured instructor-led learning by industry thought leaders such as Barry Boehm, Brian Foote, Steve McConnell, JB Rainsberger, and Linda Rising. It was during my tenure at Nortel, Qualcomm, and Cisco that I learned two challenges with scheduling talent exchanges. Corporate reorganizations and product schedules not surprisingly complicate tech transfer programs. For example, if interns are part of an organization that "disappears" due to corporate restructuring or layoffs, the interns will need to find a new corporate home. A less predictable challenge is when a "customer emergency" necessitates the cancellation of employee training. With no "live" attendees for training, the only option is to record the session without an audience to enable post event viewing.

7. SUMMARY: THE FUTURE OF WORK—ACCELERATING INNOVATION THROUGH "TECH TRANSFER"

It is hard work to effectively apply the strategies described in this experience report. To start, I recommend a three-step process. First, diagnose the organizational context to select the appropriate initial strategy. It is much easier to "start small" with a department rather than an entire business unit. Second, iterate and add other strategies as warranted by early successes. Third, track successes and learn from failure. Exploratory workshops, panels, and research reviews are good strategies to start with, while tech forums and talent exchanges generally require broader federated sponsorship.

There are two categories of stakeholders for my five strategies: R&D staff and company leaders. R&D staff participate in tech transfer activities directly, and they benefit from being part of a collaborative community. They share ideas, learn about new technologies, and adopt new best practices. Executives and product leaders benefit more indirectly: as products are catalyzed faster, at higher levels of quality and customer satisfaction.

That said, it is not easy to measure the value of tech transfer programs. The set of stakeholders is very diverse (see Table 4 for a list of the potential stakeholders for each strategy), and that makes it a challenge to find general agreement on the return on investment (ROI). My measurements of the impact and value were mostly indirect. For example, I would measure the success of tech forums by tracking data about audience size, submissions, percentage of submissions accepted for presentation, and general audience satisfaction. I also investigated what the forum presenters gained from their participation. Most of the authors of forum papers reported that they learned from the peer review process and audience feedback. Executive sponsors delighted at the downward trend of attendee "costs" and testimonials by product managers that highlighted reductions in time-to-market, improvements in product quality, and product innovations catalyzing increased company revenue.

Tech transfer activities work best when there is visible management support and stable funding. The two most important lessons I can share: Do not forget to build a broad base of executive support and do not rely on a single source for a program's budget. A program with a single executive champion makes a program more prone to "cancellation," unlike programs with federated sponsorship and budgeting.

The success of my five strategies led to industry recognition for me and the companies that implemented my strategies. For example:

- *My receipt of Cisco, Qualcomm, and Nortel awards, e.g., Nortel President's Award for R&D Partnerships*
- *American Society of Training and Development (ASTD) BEST Award for the QTech Forum*
- *Nortel, Qualcomm, and Cisco's frequent recognition by Fortune magazine as a "Best Workplace"*

Table 4. Summary of Five Strategies for Accelerating Innovation through Tech Transfer

|  | Corporate Tech Forums | Conference Panels | Exploratory Workshops | University Research Reviews | Talent Exchanges |
|---|---|---|---|---|---|
| **Purpose** | Curated proprietary program of peer reviewed presentations (crowd sourced learning) | Discussion or debate of topics by industry and university thought leaders (subject matter experts) | Facilitated discussion to share terminology and mental models, develop strategy, program road maps, and product vision | Discussions by technical staff engaged with academic research to share results of upcoming events and future trends | Academics and students obtain industry experience or industry experts share expertise with academics |
| **Participants** | R&D staff, R&D management, Executives | Panel moderator, thought leaders, audience participants | Intact project teams (R&D, marketing, HR, etc. as necessitated by project context) | New hire PhDs, R&D staff, HR (Staffing, Learning), university researchers | Academics (students, researchers), R&D staff (visiting scientists, adjunct professors) |
| **Company Stakeholders** | Participants, (R&D staff), HR (Learning), Marketing, Executives | R&D Staff Participants, Marketing | R&D Teams, Product Management | Participants (R&D Staff), Marketing, Business Development | R&D Staff HR (Talent). Product Management |
| **Accelerate Innovation** | Leverage best practices, reduce duplication, employee recognition | Tech scouting, networking, visibility for company interests | Align stakeholders, strategy roadmaps, vision statements, draft architectures | Knowledge transfer, inspire new ideas, strategic alignment | Knowledge transfer, mentoring, recruiting |
| **Challenges** | "Permission" to participate, review processes, event production, ITAR | Lack of contrarian views, long-winded questions, biased moderators | Non-actionable outcomes, poor facilitation, reluctant participants, "non-teams" | Actionable outcomes, scheduling to maximize participation, agenda setting, budgets, ITAR | Intellectual Property Rights (IPR), employee agreements, funding, ITAR |

In 2020, the COVID-19 pandemic drove companies and universities to move most of their collaborative activities to a virtual format. "Virtuality" has had positive impacts besides limiting the pandemic contagion. Many organizations have learned to initiate and sustain ongoing engagement with the tech community through university, company, or broad-spectrum conferences such as those sponsored by professional societies.

Going virtual increases the global reach and accessibility of a conference. It also reduces costs for participants and risks for organizers. Virtual options will undoubtedly multiply. All five of my tech transfer strategies will catalyze innovation, whether companies use face-to-face, virtual, or a mix of interaction modes.

I learned an important lesson early in my career: tech transfer can have unanticipated side effects. Although my initial task was to guide the incubation and adoption of "best practices," I discovered that my strategies accelerated innovation in other positive ways (see Table 4). Tech transfer through knowledge sharing helps to bring new staff members on board, captures and records corporate memory, busts corporate silos, fosters collaboration across business units, and harnesses employee enthusiasm to catalyze innovation. My co-authored workshop paper at ACM/IEEE ICSE [16], "Agile Deconstructed" keynote [17], and publication portfolio [18] illustrate how my personal journey has influenced my perspectives on tech transfer.

My experiences, captured in the five strategies described in this report, have highlighted for me the power and value of catalyzing tech transfer through knowledge sharing. My experiences have taught me to be patient, persistent, adaptive, and to approach challenges incrementally. When I initiated tech transfer strategies at BNR/Nortel and implemented virtual programs at Qualcomm, Cisco, HP, and Innoxec, I did not imagine a situation where a global pandemic would trigger the worldwide adoption of virtual tech transfer and knowledge sharing strategies. With the challenges faced by a socially distant world, these five strategies will continue to contribute to a collaborative knowledge sharing fabric which can effectively catalyze and accelerate innovation in a virtually proximate world.

## 8. ACKNOWLEDGEMENTS

Thanks go to Ken Power, Dennis Mancl, Landon Noll, Robert Crawhall, Priya Marsonia and the thousands of colleagues that participated in my conferences, forums, panels, workshops, and other tech transfer programs.

## 9. REFERENCES

[1] Henry Chesbrough: *Open Innovation*, Boston, MA. USA. Harvard Business School Press, ISBN-1-57851-837-7

[2] Steven D. Fraser, Jose M. Duran, Raymond Aubin: *Software indexing for reuse*. SMC 1989: 853-858, https://doi.org/10.1109/ICSMC.1989.71415.

[3] Mary Lynn Manns, Linda Rising: *Fearless Change: Patterns for Introducing New Ideas*, 2015: Addison-Wesley, ISBN-10: 0134395255

[4] Robert Cialdini: *Influence: The Psychology of Persuasion*, 2006: ISBN-10: 9780061241895

[5] Steven Fraser, Kent L. Beck, Grady Booch, Derek Coleman, James Coplien, Richard Helm, Kenneth S. Rubin: *How Do Teams Shape Objects? - How Do Object Shape Teams?* (Panel). OOPSLA 1994: 468-473, https://doi.org/10.1145/191080.191152.

[6] Steven Fraser, Kent L. Beck, Ward Cunningham, Ron Crocker, Martin Fowler, Linda Rising, Laurie A. Williams: *Hacker or hero? - extreme programming* today (panel session). OOPSLA Addendum 2000: 5-7, https://doi.org/10.1145/367845.367892.

[7] Steven Fraser, Kent L. Beck, Grady Booch, Larry L. Constantine, Brian Henderson-Sellers, Steve McConnell, Rebecca Wirfs-Brock, Edward Yourdon: *Echoes?: Structured design and modern software practices*. OOPSLA Companion 2005: 383-386, https://doi.org/10.1145/1094855.1094980.

[8] Steven Fraser, Frederick P. Brooks Jr., Martin Fowler, Ricardo López, Aki Namioka, Linda M. Northrop, David Lorge Parnas, Dave A. Thomas: *"No silver bullet" reloaded: Retrospective on "essence and accidents of software engineering,"* OOPSLA 2007: 1026-1030, https://doi.org/10.1145/1297846.1297973.

[9] Steven Fraser, Barry W. Boehm, Frederick P. Brooks Jr., Tom DeMarco, Tim Lister, Linda Rising, Edward Yourdon: "Retrospectives on Peopleware," *29th International Conference on Software Engineering (ICSE'07 Companion), Minneapolis, MN, USA, 2007, pp. 21-24,* http://10.1109/ICSECOMPANION.2007.61.

[10] Steven D. Fraser, Dennis Mancl, Nancy R. Mead, Mary Shaw, and Werner Wild. 2015. Software Professionalism - Is it 'Good Enough?'. In Companion Proceedings of the 2015 ACM SIGPLAN International Conference on Systems, Programming, Languages and Applications: Software for Humanity (SPLASH Companion 2015). Association for Computing Machinery, New York, NY, USA, 60–62. https://doi.org/10.1145/2814189.2818718

[11] S. D. Fraser and C. S. Saunders, "Enhanced reuse with group decision support systems," [1993] Proceedings Advances in Software Reuse, Lucca, Italy, 1993, pp. 168-175, http://10.1109/ASR.1993.291706.

[12] Managing Technical Debt in Software Engineering, Schloss Dagstuhl, 2016: https://www.dagstuhl.de/en/program/calendar/semhp/?semnr=16162

[13] Steven D. Fraser, 2015. *Reflections on software engineering research collaborations,* Proc. 2nd Intl. Workshop on Software Engineering Research and Industrial Practice (SER&IP 2015), ACM, 2015, pp. 5-10, http://10.1109/SERIP.2015.10.

[14] Steven D. Fraser and D. Mancl, 2016. *Strategies for building successful company-university research collaborations*, Proc. 3rd Intl. Workshop on Software Engineering Research and Industrial Practice (SER&IP 2016), Austin, TX, 2016, pp. 10-15, https://10.1145/2897022.2897025.

[15] S. Fraser and D. Mancl, "Innovation through Collaboration: Company-University Partnership Strategies," 2017 IEEE/ACM 4th International Workshop on Software Engineering Research and Industrial Practice (SER&IP), Buenos Aires, Argentina, 2017, pp. 17-23, https://10.1109/SER-IP.2017..3.

[16] S. Fraser and D. Mancl, "Exploring the Dimensions of University-Company Collaborations: Research, Talent, and Beyond," *2021 IEEE/ACM 8th International Workshop on Software Engineering Research and Industrial Practice (SER&IP),* Madrid, Spain, 2021, pp. 57-64, https://10.1109/SER-IP52554.2021.00017.

[17] Steven Fraser: *Software "Best" Practices: Agile Deconstructed*. PROFES 2009: 8-13, https://doi.org/10.1007/978-3-642-02152-7_2

[18] https://dblp.org/pid/75/1446.html